\documentclass[review]{elsarticle}

\usepackage{hyperref} % lineno for numbering
\usepackage{amsmath,amssymb,amsthm, bm}
\usepackage{algorithm}
\usepackage{algorithmic}
\usepackage{relsize}
\usepackage{amsmath,amssymb,amsthm, bm}
\usepackage{graphicx}
\usepackage{subfig}

\journal{arXiV} %Biomedical Physics \& Engineering Express
\newcommand{\etal}{\textit{et al. }}
\begin{document}

\begin{frontmatter}

\title{EOG Artifact Removal from Single and Multi-channel EEG Recordings through the combination of Long Short-Term Memory Networks and Independent Component Analysis}

%Commented for double anonymous review
\author[1,3]{\corref{mycorrespondingauthor} Behrad TaghiBeyglou\footnote{$^1$ B. TaghiBeyglou was with School of Electrical Engineering, Sharif University of Technology, Tehran, Iran.}}
\cortext[mycorrespondingauthor]{Corresponding author}
\ead{behrad.taghibeyglou@mail.utoronto.ca}

\author[2]{Fatemeh Bagheri\footnote{$^{2}$ F. Bagheri was with Biomedical Engineering Department, Amirkabir University of Technology, Tehran, Iran.}}

\address[1]{Institute of Biomedical Engineering, University of Toronto, Toronto, Canada}
\address[3]{KITE- Toronto Rehabilitation Institute, University Health Network, Toronto, Canada}
\address[2]{Deparment of Medical Biophysics, University of Toronto, Toronto, Canada}
% %% or include affiliations in footnotes:

\begin{abstract}
\textbf{Introduction}: Electroencephalogram (EEG) signals have gained significant popularity in various applications due to their rich information content. However, these signals are prone to contamination from various sources of artifacts, notably the electrooculogram (EOG) artifacts caused by eye movements. The most effective approach to mitigate EOG artifacts involves recording EOG signals simultaneously with EEG and employing blind source separation techniques, such as independent component analysis (ICA). Nevertheless, the availability of EOG recordings is not always feasible, particularly in pre-recorded datasets.
\textbf{Objective}: In this paper, we present a novel methodology that combines a long short-term memory (LSTM)-based neural network with ICA to address the challenge of EOG artifact removal from contaminated EEG signals. 
\textbf{Approach}: Our approach aims to accomplish two primary objectives: 1) estimate the horizontal and vertical EOG signals from the contaminated EEG data, and 2) employ ICA to eliminate the estimated EOG signals from the EEG, thereby producing an artifact-free EEG signal.
\textbf{Main results}: To evaluate the performance of our proposed method, we conducted experiments on a publicly available dataset comprising recordings from 27 participants. We employed well-established metrics such as mean squared error, mean absolute error, and mean error to assess the quality of our artifact removal technique. \textbf{Significance}: Furthermore, we compared the performance of our approach with two state-of-the-art deep learning-based methods reported in the literature, demonstrating the superior performance of our proposed methodology.
\end{abstract}


\end{frontmatter}

%\linenumbers

\section{Introduction}
Brain-computer interfaces (BCIs) are systems that transform raw brain activities of humans into interpretable information that can control other devices or reflect the state of rehabilitation or disease status \cite{nguyen2012eog,ramadan2017brain,qiu2016improved}. One of the most prominent modalities to record brain activation is the Electroencephalogram (EEG), which primarily records scalp voltages non-invasively \cite{yang2018automatic,ferdowsi2018multi,cudlenco2020reading}. These voltages represent the superposition of many neural activities and contribute to different tasks \cite{grech2008review}. However, as EEG recording is performed on the scalp, physiological activities other than brain signals can interfere and introduce artifacts \cite{croft2000removal}. These unwanted interferences, known as artifacts, need to be removed to ensure accurate EEG analysis \cite{mehdizadehfar2020eeg,shooshtari2006removing}.

One of the most significant physiological artifacts is the Electrooculogram (EOG), which directly relates to eye movements and blinks. Due to the close proximity between the eyes and the scalp, the effect of the EOG artifact is undeniable \cite{scott2002electroolfactogram,karimi2013automatic}. Eye blinking and movements generate spike-like signal waveforms, with peak amplitudes reaching up to $800\mu$V and occurring within a short period of $200$--$400$ ms \cite{hagemann2001effects,yang2016removal}. Additionally, EOG artifacts often overlap with EEG signals in the frequency domain, particularly at low frequencies, as well as in the time domain \cite{nguyen2012eog,fatourechi2007emg}. Since the extraction of features and information from EEG signals for classification and analysis requires a clean signal, the removal of EOG artifacts is essential for subsequent steps \cite{mehdizadehfar2020eeg,shooshtari2006removing}.

Since the introduction of the EEG signal by Berger \cite{berger1931elektrenkephalogramm}, researchers have faced the challenge of EOG artifacts \cite{imaizumi1967origin}. With the development of mathematical tools, significant attention has been devoted to removing EOG artifacts from EEG signals \cite{elbert1985removal,gevins1975automated,girton1973simple}. Most of these methods rely on blind source separation (BSS), particularly independent component analysis (ICA). However, these techniques are most effective when EOG signals are available alongside EEG recordings. In cases where EOG signals are not available, which is common in pre-recorded datasets, researchers have sought methods to directly estimate EOG signals from contaminated or artifact-included EEG signals. To provide a comprehensive review of the literature, we categorize different EOG artifact removal methods into three categories: Single Channel, Multi-channel EEG recording, and methods effective for both scenarios, as some approaches claim to remove EOG artifacts using multi-channel EEG recordings, while others rely on a single channel.

\section{Literature review}
\subsection{Single Channel} 
In 1973, Girton \etal introduced a simple hardware-based method for online removal of EOG artifacts from EEG recordings \cite{girton1973simple}. Their approach involved using a circuit to record EOG signals from two electrodes (related to Horizontal and Vertical movements of the eyes) and another circuit to record EEG signals. By utilizing a scaling and subtraction circuit, they successfully removed Horizontal EOG (HEOG) and Vertical EOG (VEOG) from the raw EEG. However, this method relied on the availability of hardware and separate EOG electrodes, making it limited in its applicability. Nonetheless, it served as the foundation for subsequent advancements in the field of EOG artifact removal.

Nearly 20 years later, in 2004, He \etal proposed an adaptive filtering-based method for EOG removal \cite{he2004removal}. Their approach required an EOG electrode, which, when combined with another EEG electrode, facilitated the construction of an adaptive filter with a filter order of M=3. While effective, the main limitation of this method was its dependence on the presence of a separate EOG recording.

In 2014, Hu \etal  introduced a novel approach combining an adaptive neural fuzzy inference system (ANFIS) and a functional link neural network (FLNN) to remove EOG and electromyogram (EMG) artifacts from EEG signals \cite{hu2015removal}. The method involved an adaptive filtering algorithm that adjusted the parameters of the fuzzy inference and neural network. Although successful in artifact removal, this technique still relied on the availability of raw artifact data in the early stages of the removal process.

Another approach proposed in 2014 by Maddirala \etal utilized singular spectrum analysis (SSA) and adaptive noise canceler (ANC) to remove EOG artifacts from contaminated EEG signals \cite{maddirala2016removal}. The technique involved grouping the SSA components to construct an EOG reference signal for ANC. Using this estimated reference signal, an adaptive filter was employed to remove the EOG artifact. Performance evaluation using RRMSE (relative root mean square error) and MAE (Mean Absolute Error) metrics demonstrated the effectiveness of the proposed algorithms. Notably, this method only required EEG electrodes and did not rely on the presence of dedicated EOG channels, enhancing its reliability in situations where EOG electrodes were not available.

In 2020, Noorbasha \etal presented a research study introducing the overlap segmented adaptive singular spectrum analysis (Ov-ASSA) combined with adaptive noise canceler (ANC) technique for EOG artifact removal \cite{noorbasha2020removal}. In this approach, the first one or two reconstructed components of the Ov-SSA technique were adaptively grouped and used as a reference EOG signal for ANC in a single-channel EEG recording system. The algorithm's performance was evaluated using simulated and real data, with RRMSE and MAE metrics employed as performance measures.

\subsection{Multi Channel}
In 2000, Jung \etal presented a method for removing various artifacts from EEG records using BSS through ICA \cite{jung2000removing}. Their study demonstrated the effectiveness of ICA in detecting, separating, and removing contamination from different artifactual sources in EEG records. However, their method relied on the availability of EOG electrodes during EEG recording, which is a significant limitation. Nonetheless, their contribution in the field of artifact removal has been influential and serves as inspiration for our proposed method.

Shahabi \etal introduced a method for removing eye blink artifacts from EEG signals using EOG reference electrodes \cite{shahabi2012eeg}. They employed an autoregressive (AR) process to model the EEG activity and an output-error model for eye blinks. By utilizing a Kalman filter, they estimated the actual EEG by merging the two models. The performance of their method was evaluated using the BCI competition 2008, dataset II-a, which is also one of the datasets used for performance evaluation in our study.

Nguyen \etal published a framework in 2012 that combined artificial neural networks and wavelet transformation for EEG artifact removal \cite{nguyen2012eog}. Their algorithm utilized the universal approximation characteristics of neural networks and the time-frequency properties of the wavelet transform. The neural network was trained on a simulated dataset with known ground truths. Notably, their framework did not require EOG electrodes during the artifact removal procedure, distinguishing it from many other EEG artifact removal algorithms. They also compared their results with the ICA method using a time-consuming procedure for artifact rejection on simulated and real datasets.

Zeng \etal proposed a mixed-method approach in which they used singular spectrum analysis (SSA) as a blind source separation technique to extract different components, including artifactual and neural components, from EEG signals \cite{elbert1985removal}. They then employed empirical mode decomposition (EMD) to denoise the components affected by EOG artifacts. The artifactual components were projected back to the electrode space and subtracted from the EEG signals to obtain clean EEG. The experimental results on artificially contaminated EEG data and publicly available real EEG data were reported. However, their method still required EOG recording electrodes during EEG recording.

In 2015, Yang \etal proposed the ICA-based multivariate empirical mode decomposition (IMEMD) method for removing EOG artifacts from multichannel EEG signals \cite{wang2015removal}. They decomposed the EEG signals using multivariate empirical mode decomposition (MEMD) into multiple multivariate intrinsic mode functions (MIMFs). The artifactual components were extracted by reconstructing the MIMFs corresponding to EOG artifacts. By performing ICA on the contaminated signals, the EOG-related independent components (ICs) were identified and removed. Finally, the denoised EEG signals were reconstructed by applying the inverse transform of ICA and MEMD. They evaluated the performance of their method using signal-to-noise ratio (SNR) and mean squared error (MSE). However, this technique still required EOG electrodes during EEG recording.
\subsection{Both single channel and multi channel}
In 2016, Yang \etal proposed a method that combines a sparse autoencoder (SAE) with recursive least square adaptive (RLS) filtering to remove EOG artifacts without the need for reference electrodes \cite{yang2016removal}. During the offline step, the SAE model learns information from EOG signals, and in the online stage, the trained SAE model is used to extract preliminary EOG artifacts from the raw EEG signal. The recursive least square adaptive filter then uses the identified EOG artifacts as a reference signal to remove interference without parallel EOG recordings. Their method was evaluated using a classification accuracy metric.

In 2018, Yang \etal published another research in which they utilized a deep learning network (DLN) to remove EOG artifacts from EEG signals \cite{yang2018automatic}. The DLN consisted of a multi-layer perceptron (MLP) with an SAE-based approach to estimate non-artifactual EEG from the raw EEG signal. The proposed method involved two stages. In the offline stage, training samples without artifacts were used to train the DLN, enabling it to reconstruct denoised EEG signals and learn the high-order statistical moments of the EEG. In the online stage, the learned DLN was applied as a filter to remove EOG artifacts from contaminated EEG signals.

Considering the advantages and limitations of previous methods, our study aims to present a method for EOG artifact removal without requiring an EOG reference electrode during the procedure, applicable to both single-channel and multi-channel EEG signals. Additionally, we aim to estimate both horizontal (HEOG) and vertical (VEOG) EOG signals and remove these artifactual components from the EEG signal. Our approach is based on long short-term memory (LSTM) neural networks and ICA. The study involves offline and online stages. In the offline stage, the LSTM network learns features of EOG signals from EEG recordings after signal normalization. In the online stage, the EEG signals are processed by the LSTM network to estimate the EOG signal, which is then used as the EOG reference electrode. By performing ICA, all components, including artifactual and non-artifactual ones, are extracted. Components similar to the estimated EOG signal are identified as artifactual and removed from the sources. Finally, the denoised EEG signal is estimated by back-projecting to the electrode subspace. 

Throughout this paper, vectors are represented by bold small letters, e.g., $\mathbf{x}$, matrices by capital bolded letters, e.g., $\mathbf{X}$, and scalars by normal italic fonts, e.g., $x$. Element-wise multiplication is denoted by "$\circ$", matrix multiplication by "$\times$", and concatenation by "$[,]$". The $l^2$-norm of a vector is denoted as "$|\lvert . |\lvert$", and the transpose and inverse of a matrix are denoted by superscript $\top$ and $-1$, respectively.

The remainder of the paper is organized as follows: Section \ref{Method} introduces the proposed method and the employed dataset and provides a detailed description of the methodology. Section \ref{results} presents the experimental setup and results. Finally, Sections \ref{discussion} and \ref{conclusion} conclude the paper with a discussion of the findings.

\section{Method}\label{Method}
\subsection{LSTM}
LSTM networks were first introduced by Hochreiter et al. in 1997 as a specific type of recurrent neural network capable of learning long-term dependencies \cite{hochreiter1997long}. These networks were designed to address the challenge of capturing and retaining information over extended periods. Notably, LSTM networks excel at memorizing information for prolonged durations by leveraging their unique architecture \cite{gensler2016deep,hafezi2020sleep}. The structure of LSTM networks enables them to learn and retain information effectively, which distinguishes them from other recurrent neural networks. In general, recurrent neural networks consist of a repetitive sequence of simple neural network modules or units.

\begin{figure}[ht!]
    \centering
    \includegraphics[width=\textwidth]{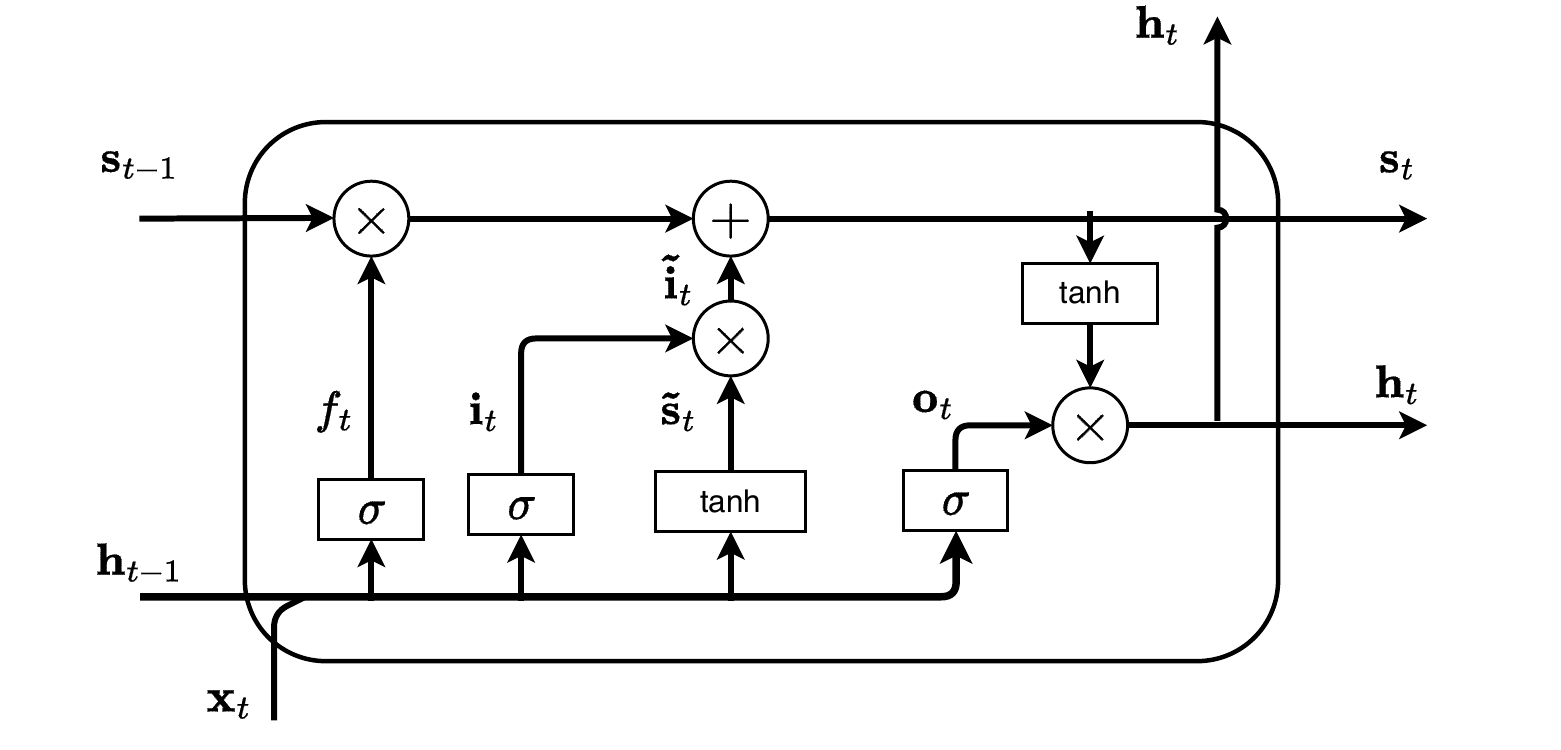}
    \caption{Diagram of an LSTM cell. }
    \label{fig:LSTM}
\end{figure}

The core component of LSTM networks is the LSTM cell, illustrated in Figure \ref{fig:LSTM}. The annotations used in the figure and in this section are inspired by \cite{wang2018lstm}.

In this context, $\mathbf{x}_t$ represents the input vector at the current time step, while $\mathbf{s}_t$ and $\mathbf{h}_t$ are the cell state and the output of the hidden layer at time $t$, respectively. Both $\mathbf{s}_t$ and $\mathbf{h}_t$ are initialized as zeros (i.e., $\mathbf{s}_0 = \mathbf{h}_0 = \mathbf{0}$). The LSTM cell computes $\mathbf{s}_t$ and $\mathbf{h}_t$ as follows:
\begin{align}
\mathbf{s}_t &= \mathbf{s}_{t-1} \circ \mathbf{f}_t + \mathbf{\tilde{i}}_t, \\
\mathbf{h}_t &= \mathbf{o}_t \circ \textrm{tanh}(\mathbf{s}_t),
\end{align}
where $\mathbf{f}_t$, $\mathbf{i}_t$, and $\mathbf{o}_t$ denote the outputs of the forget, input, and output gates, respectively.

The forget gate determines the amount of information from $\mathbf{s}_{t-1}$ that should be discarded. Its output is computed as:
\begin{equation}
\mathbf{f}_t = \sigma(\mathbf{W}_f\times[\mathbf{h}_{t-1},\mathbf{x}_t] + \mathbf{b}_f),
\end{equation}
where $\mathbf{W}_f$ and $\mathbf{b}_f$ are trainable coefficients, and $\sigma(\cdot)$ represents the sigmoid function defined as:
\begin{equation}
\sigma(z) = \frac{1}{1 + e^{-z}}.
\end{equation}

The input gate determines the amount of information from $\mathbf{x}_t$ and $\mathbf{h}_{t-1}$ that should be retained in the current cell state $\mathbf{s}_t$. Its output is computed as:
\begin{equation}
\mathbf{i}_t = \mathbf{\tilde{i}}_t \circ \mathbf{\tilde{s}}_t,
\end{equation}
where
\begin{align}
\mathbf{\tilde{i}}_t &= \sigma(\mathbf{W}_i\times [\mathbf{h}_{t-1},\mathbf{x}_t] + \mathbf{b}_i), \\
\mathbf{\tilde{s}}_t &= \textrm{tanh}(\mathbf{W}_s\times [\mathbf{h}_{t-1},\mathbf{x}_t] + \mathbf{b}_s),
\end{align}
with $\mathbf{W}_i$, $\mathbf{W}_s$, $\mathbf{b}_i$, and $\mathbf{b}_s$ representing weighting matrices and biases, respectively. The function $\textrm{tanh}(\cdot)$ denotes the hyperbolic tangent function defined as:
\begin{equation}
\textrm{tanh}(z) = \frac{e^z - e^{-z}}{e^z + e^{-z}}.
\end{equation}

The output gate determines the amount of information from $\mathbf{s}_t$ that should be included in the output of the hidden layer $\mathbf{h}_t$. Its output is computed as:
\begin{equation}
\mathbf{o}_t = \sigma(\mathbf{W}_o\times [\mathbf{h}_{t-1},\mathbf{x}_{t}]+ \mathbf{b}_o),
\end{equation}
where $\mathbf{W}_o$ and $\mathbf{b}_o$ represent the corresponding weighting matrix and bias vector.

In this study, we employ a deep LSTM network consisting of four layers, each with 64 hidden units and accompanied by dropout rates of 0.1, 0.3, 0.3, and 0.1, respectively. The last layer is connected to a fully connected network with an output size of 2, representing the VEOG and HEOG signals.
% The main element of LSTMs is the cell state, a horizontal line at the top of the figure.
% The state cell can be thought of as a conveyor belt that moves from the beginning to the end of a sequence or chain with minor linear interactions (i.e., its structure is straightforward, and there are few changes in it).
% LSTM can add new information to the state cell or delete its information; This is done by precise structures called gates.
% Gates are a way for information to enter voluntarily. They consist of a layer of a sigmoid neural network with a point-to-point multiplier.
% The output of the sigmoid layer is a number between zero and one, indicating how much of the surface should be sent to the output. Zero value means no information should be sent to the output, while value one means all input to the output!
% LSTM has three similar gates to control the value of state cells.

\subsection{ICA}
EEG signals recorded from scalp electrodes represent a combination of neural activations occurring in different areas of the brain \cite{ungureanu2004independent, jung1998extended}. The scalp electrode's location determines the weights assigned to each neural activity. The primary objective is to extract the dominant activity from these combined activations to gain insights into the underlying brain sources. While invasive recording allows for direct measurement of specific neural activity, it requires surgery and is generally not feasible or desirable. Instead, scalp EEG recordings provide a convenient means to capture neural activity, albeit with a mixture of different activations weighted differently across electrode channels.

If we had knowledge of these weights, we could compute the potentials in the brain sources using a sufficient number of electrodes \cite{ungureanu2004independent}. However, the exact weights and the contribution of each neural activity are typically unknown. Consequently, we rely on BSS methods to overcome this challenge.

ICA is a widely recognized BSS algorithm \cite{hyvarinen2000independent} and aims to extract sources by maximizing non-gaussianity and minimizing mutual information \cite{taymourtash2015independent}. It assumes that the observed signals are linear mixtures of these independent sources. Various implementations of ICA, such as Infomax, SOBI, JADE, FastICA, and kernel-independent component analysis, have been proposed \cite{matic2009comparison}. ICA is well-suited for EEG signals due to two key assumptions about the recordings:
\begin{enumerate}
    \item The observations from different electrodes on the scalp correspond to neural activations from distinct brain areas, assuming independence between areas.
    \item The generation and recording speed of EEG signals are much slower than the speed of electromagnetic wave propagation, resulting in negligible delays between the origin of neural activity and the recording electrodes.
\end{enumerate}
These assumptions allow ICA to effectively separate the contributions of different sources from scalp EEG recordings, providing valuable insights into the underlying neural activity.

Consider $\mathbf{x}j$ as the vector signal of the j-th channel, and $\mathbf{a}_{1}$ to $\mathbf{a}_{N_s}$ as the projection vectors. Additionally, let $\mathbf{s}_1$ to $\mathbf{s}_{N_s}$ represent the vector sources to be estimated. The number of channels, sources, and time samples are denoted as $N_c$, $N_s$, and $T$, respectively. The overall problem can be expressed as follows (Equation \eqref{eq1}):

\begin{equation}
\label{eq1}
\mathbf{x}_j = a_{j,1}\mathbf{s}_1 + \cdots + a_{j,N_s}\mathbf{s}_{N_s}
\end{equation}

To generalize Equation \eqref{eq1} to all channels and sources using the projection vectors, we can write Equation \eqref{eq2}. Each vector is defined as follows: $\mathbf{x}_j = [x_{j,1},\cdots,x_{j,T}] , j=1,...,N_c$, $\mathbf{a}_i = [a_{i,1},\cdots,a_{i,N_c}]^T, i=1,...,N_s$, and $\mathbf{s}_i = [s_{i,1},\cdots,s_{i,T}] , i=1,...,N_s$.

\begin{equation}
\label{eq2}
\begin{bmatrix}
\mathbf{x}_{1} \\
\vdots\\
\mathbf{x}_{N_c}\\
\end{bmatrix} = \begin{bmatrix}
\mathbf{a}_1& \cdots& \mathbf{a}_{N_s}
\end{bmatrix} \times \begin{bmatrix}
\mathbf{s}_{1} \\
\vdots\\
\mathbf{s}_{N_s}\\
\end{bmatrix}
\qquad \Rightarrow \qquad \mathbf{X} = \mathbf{A} \times \mathbf{S} = \sum_{i = 1}^{N_s} \mathbf{a}_i^T\mathbf{s}_i
\end{equation}

The above equation represents the problem of finding the sources matrix ($\mathbf{S}$) and the mixing matrix ($\mathbf{A}$). To reduce the dependency on the center and variance of the observations, the observed data ($\mathbf{X}$) should be whitened. The centering and whitening procedure can be performed using Algorithm \ref{alg1}.

\begin{algorithm}[ht!]
	\caption{Centering and Whitening algorithm}
	\label{alg1}
	\begin{algorithmic}[1]	
		
		\REQUIRE $\mathbf{X}$ as the observation\\
		\ENSURE $\mathbf{\tilde{X}}$ as the whitened matrix\\[5pt]
		\STATE $\mathbf{\hat{X}} \leftarrow \mathbf{X}-\mathbb{E}\{\mathbf{X}\}$
		\STATE Eigenvalue decomposition of the covariance matrix of $\mathbf{\hat{X}}$ $\rightarrow$ $\mathbf{V}\mathbf{D}\mathbf{V}^T = \mathbb{E}\{\mathbf{\hat{X}}\mathbf{\hat{X}}^T\}$
		\STATE $\mathbf{P}\leftarrow\mathbf{D}^{-\frac{1}{2}}\mathbf{V}^T$
		\STATE $\mathbf{\tilde{X}}\leftarrow\mathbf{P}\mathbf{\hat{X}}$
		\RETURN $\mathbf{\tilde{X}}$
	\end{algorithmic}
\end{algorithm}

After performing the whitening procedure, the next step is to solve the problem of finding the sources and mixing matrix arrays. In this study, we utilized FastICA, which is based on a fixed-point iteration scheme that aims to maximize the non-Gaussianity of $\mathbf{W}^T\mathbf{X}$ \cite{hyvarinen2000independent}. The iteration process is achieved using the Newton iterative method \cite{hyvarinen1999fast}. To measure non-Gaussianity, FastICA employs a non-quadratic nonlinear function $f(u)$, its first derivative $g(u)$, and its second derivative $g^\prime(u)$. These functions are defined in Equation \eqref{eq3}. The algorithm for extracting multiple components using the FastICA method is outlined in Algorithm \ref{alg2}, where $\mathbf{1}$ represents the column vector of size $T \times 1$ filled with the value 1.

\begin{equation}
\label{eq3}
f(u) = -e^{-\frac{u^2}{2}} \xrightarrow{\frac{\partial}{\partial u}} g(u) = ue^{-\frac{u^2}{2}}\xrightarrow{\frac{\partial}{\partial u}} g^\prime(u) = (1-u^2)e^{-\frac{u^2}{2}}
\end{equation}
\begin{algorithm}[ht!]
	\caption{FastICA algorithm}
	\label{alg2}
	\begin{algorithmic}[1]	
		
		\REQUIRE $\mathbf{\tilde{X}} \in {\rm I\!R}^{N_c\times T}$  as the whitened observation, and $N_s$ which is the number of proposed components ($\leq N_s$)\\
		\ENSURE $\mathbf{W} \in {\rm I\!R}^{N_c\times N_s}$ as the unmixing matrix, $\mathbf{S} \in {\rm I\!R}^{N_s\times T}$ independent component matrix  \\[5pt]
		\STATE Initialize $\mathbf{W}$ with random values ($\sim \mathcal{N}(\mu=0,\,\sigma^{2}=1)$)
		\FOR{$i$ in $1$ to $N_s$}
		\WHILE{$\mathbf{w}_i$ changes}
		\STATE $\mathbf{w}_i \leftarrow \frac{1}{T} \mathbf{\tilde{X}}g(\mathbf{w}_i^T\mathbf{\tilde{X}})^T - \frac{1}{T}g^\prime(\mathbf{w}_i^T\mathbf{\tilde{X}})\mathbf{1}\mathbf{w}_i$
		\STATE $\mathbf{w}_i \leftarrow \mathbf{w}_i - \mathlarger{\sum}_{j=1}^{i-1} (\mathbf{w}_i^T\mathbf{w}_j)\mathbf{w}_j$
		\STATE $\mathbf{w}_i \leftarrow \frac{\mathbf{w}_i}{\left\lVert\mathbf{w}_i\right\rVert}$
		\ENDWHILE
		\ENDFOR
		\STATE $\mathbf{W} \leftarrow [\mathbf{w}_1,\mathbf{w}_2,\cdots,\mathbf{w}_{N_s}]$
		\STATE $\mathbf{S} \leftarrow \mathbf{W}^T\mathbf{\tilde{X}}$
		\RETURN $\mathbf{W},\mathbf{S}$
	\end{algorithmic}
\end{algorithm}
\subsection{Combining LSTM and ICA}
This paper introduces two different approaches: a single-channel approach and a multi-channel approach. Each approach consists of two stages: an offline stage and an online stage. In the offline stage, a Deep LSTM network is trained to learn the underlying time series, sequences, and features in both contaminated EEG signals and EOG recordings. In the online stage, the estimated EOGs are treated as external EOG channels and are combined with other EEG recording electrodes. Subsequently, ICA is applied to extract both clean and artifactual sources. Finally, the artifactual sources are removed during back projection, resulting in a cleaned recording. The estimated EOG recordings are also provided as part of the outputs.
\subsubsection{Training LSTM network in offline stage}
To train the proposed neural network, the number of input and output channels needs to be specified. Since our network is designed for both single-channel and multi-channel EOG artifact removal, we consider both single and multiple inputs to account for the removal of EOG artifacts in each dataset.

Our framework is also flexible in terms of the number of input samples. This means that when we feed the network with varying sample sizes, the output can be generated without any limitations on data segmentation or other sampling techniques. During the training scheme, we used a batch size of 250 samples per iteration at each epoch. However, the segment size is not fixed, and during testing, we can feed the network with different input sample sizes.

For the loss function during the training of the Deep LSTM network, we utilized Mean Squared Error (MSE). This loss function, as formulated in Equation \eqref{eq4}, is chosen due to its differentiability and sensitivity to outliers.

\begin{equation}
\label{eq4}
MSE = \frac{1}{N}\sum_{i=1}^{N} (\hat{y}_i - y_i)^2
\end{equation}

To optimize the Deep LSTM network weights and parameters, we employed the Adam algorithm \cite{kingma2014adam}. This algorithm utilizes first-order gradient optimization based on adaptive momentum estimation. It is easy to implement, computationally efficient, requires minimal memory, is invariant to diagonal rescaling of gradients, and is well-suited for problems involving large datasets and/or parameters \cite{kingma2014adam}.

We trained the network for 50 epochs and implemented an early stopping scheme with a patience of 2 epochs based on validation loss  to avoid over-fitting on training data \cite{caruana2001overfitting}.

Once the training is completed, the network is ready to estimate EOG recordings (with 2 channels for VEOG and HEOG) and the next step involves using the ICA algorithm to extract non-artifactual sources and reconstruct the clean EEG recordings.

\subsection{Investigating ICA for EOG artifact removal in online stage} 
The estimated EOG recordings are combined with the normalized EEG recordings (without EOG references) to form the EOG electrodes. The resulting observation is then subjected to whitening, followed by the application of the FastICA algorithm to extract sources ($\mathbf{S}$) and the unmixing matrix ($\mathbf{W}$). At this stage, it is necessary to identify and remove EOG artifactual sources in subsequent steps. To accomplish this, we compute the absolute correlation index (due to the scale uncertainty in the ICA method) between each source and the estimated EOG recordings from the previous step. Sources with a high correlation (above 0.8) are considered potential artifactual sources, and their corresponding columns in the mixing matrix ($\mathbf{A}$) are set to zero to eliminate their effect during the reconstruction stage. The complete procedure is outlined in Algorithm \ref{alg4}. It is important to note that while our formulation assumes two EOG estimations (generated by the Deep LSTM network, mainly related to VEOG and HEOG), our framework can accommodate different number of EOG channels. The ";" command represents vertical concatenation, and \texttt{corrcoeff} refers to the temporal correlation measurement function.

\begin{algorithm}[ht!]
	\caption{Application of ICA in EOG artifact removal}
	\label{alg4}
	\begin{algorithmic}[1]	
		
		\REQUIRE $\mathbf{X}_N$ as the normalized EEG recording, $\mathbf{\hat{Y}}$ as the estimated EOG from Deep LSTM network, and $N_c$, $N_E$ as the number of EEG and EOG channels, respectively\\
		\ENSURE $\mathbf{X}_r$ as the reconstructed EEG\\[5pt]
		
		\STATE $\mathbf{Z} \leftarrow [\mathbf{X_N};\mathbf{\hat{Y}}]$
		\STATE $\mathbf{Z} \leftarrow \texttt{Whitening}(\mathbf{Z})$ using Algorithm \ref{alg1}
		\STATE $N \leftarrow N_c + N_E$
		\STATE $[\mathbf{S},\mathbf{W}] \leftarrow \texttt{FastICA}(\mathbf{Z},N)$ using Algorithm \ref{alg2}
		
		\FOR{$i$ in $1$ to $N$}
		\STATE $corr_1 = |\texttt{corrcoeff}\left(\mathbf{\hat{Y}}(1,:),\mathbf{S}(i,:)\right)|$
		\STATE $corr_2 = |\texttt{corrcoeff}\left(\mathbf{\hat{Y}}(2,:),\mathbf{S}(i,:)\right)|$
		\ENDFOR
		\STATE $id_1 = \texttt{find}\left(corr_1\geq0.8\right)$
		\STATE $id_2 = \texttt{find}\left(corr_2\geq0.8\right)$
		\STATE $\mathbf{A} \leftarrow \mathbf{W}^{-1}$
		\STATE $\mathbf{A}(:,id_1) \leftarrow \mathbf{0}$
		\STATE $\mathbf{A}(:,id_2) \leftarrow \mathbf{0}$
		\STATE $\mathbf{X_r} \leftarrow \mathbf{A}\mathbf{S}$
		\STATE $\mathbf{X}_r \leftarrow \mathbf{X}_r(1:N_c,:)$
		\RETURN $\mathbf{X_r}$
	\end{algorithmic}
\end{algorithm}

\begin{figure}
    \centering
    \includegraphics[width=\textwidth]{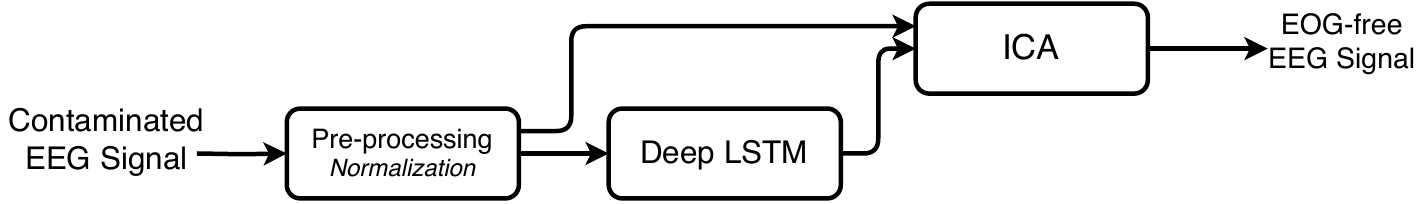}
    \caption{The flowchart of the proposed model.}
    \label{fig:diagram}
\end{figure}

To remove EOG artifacts from single-channel EEG recordings, Algorithm \ref{alg4} can still be applied with a slight modification in the notation. Instead of representing the observation as a matrix $\mathbf{X}$, it is now represented as a vector $\mathbf{x}$. The overall process of model development is illustrated in Figure \ref{fig:diagram}, showcasing the different steps involved in EOG artifact removal.

\subsection{Evaluation metrics}
The performance of the EOG estimation and the effectiveness of the algorithm in reconstructing clean EEG signal are evaluated using three metrics: Mean Squared Error (MSE), Mean Error (ME), and Mean Absolute Error (MAE). Assuming the non-contaminated original EEG signal is denoted as $\mathbf{y} \in \mathbb{R}^{N}$ and the estimated artifact-free signal is $\hat{\mathbf{y}}$, the metrics are defined as follows:

\begin{align}
MSE &= \frac{1}{N} \sum_{i=1}^N (y_i - \hat{y}i)^2 \\
MAE &= \frac{1}{N} \sum{i=1}^N \lvert y_i - \hat{y}i \rvert \\
ME &= \frac{1}{N} \sum{i=1}^N (y_i - \hat{y}_i).
\end{align}
\subsection{Dataset}
One of the crucial aspects in developing an artifact removal algorithm is the availability of suitable data for evaluating its effectiveness. Since real datasets often lack a direct benchmark for artifact removal and only provide measurements of accuracy or performance in specific tasks, it becomes necessary to generate synthetic data that includes both EOG recordings and pure EEG recordings. In this project, we utilized a semi-simulated EEG/EOG dataset consisting of fifty-four datasets recorded from twenty-seven healthy subjects (age: $27.17\pm5.2$).

\begin{figure}[!ht]
	\centering
	\includegraphics[width=\linewidth]{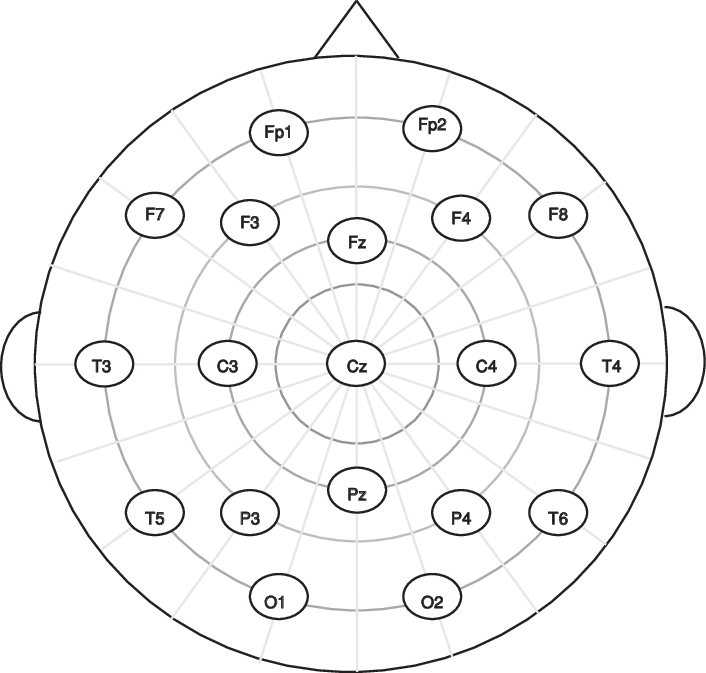}
	\caption{Channel location for the dataset used in this study.}
	\label{fig:kladosloc}
\end{figure}

Each dataset includes nineteen EEG electrodes (FP1, FP2, F3, F4, C3, C4, P3, P4, O1, O2, F7, F8, T3, T4, T5, T6, Fz, Cz, Pz) placed according to the 10-20 International System (as shown in Fig.\ref{fig:kladosloc}), along with four EOG electrodes used to record VEOG and HEOG eye movements in a bipolar scheme \cite{klados2016semi}.

The recorded datasets consist of 30 seconds of closed eyes without any eye movements to capture clean and pure EEG signals. The signals were sampled at a frequency of 200 Hz and filtered using a bandpass filter (0.5 to 40 Hz). In parallel, EOG recordings were acquired from the same subjects and during the same time duration. These EOG signals were then filtered using a 0.5 to 5 Hz bandpass filter and added to the pure EEG signals using Equation \eqref{eq5} to contaminate them.

The parameters $a$ and $b$ in Equation \eqref{eq5} represent subject- and dataset-specific coefficients that were computed using linear regression to introduce blink and eye movement artifacts into the pure EOG signals.

\begin{equation}
\label{eq5}
\mathbf{X}_{\textrm{Contaminated}} = \mathbf{X}_{\textrm{Pure}} + a\mathbf{X}_{\textrm{VEOG}} + b\mathbf{X}_{\textrm{HEOG}}
\end{equation}
% TODO: \usepackage{graphicx} required
\subsection{Pre-processing}
To mitigate the impact of varying signal values and potential outliers in EEG and EOG signals during the optimization algorithm's training phase, it is crucial to normalize these signals. This normalization involves applying a zero mean unit variance normalization to both the EEG and EOG channels. Additionally, the means and standard deviations of the signals are stored in vectors for reconstruction purposes after network training. The normalization procedure is outlined in Algorithm \ref{alg3}. 
\begin{algorithm}[ht!]
	\caption{Normalization Algorithm}
	\label{alg3}
	\begin{algorithmic}[1]	
		
		\REQUIRE $\mathbf{X} \in {\rm I\!R}^{N_c\times T}$  as input EEG observation, $\mathbf{Y} \in {\rm I\!R}^{N_E\times T}$ as the output EOG recording\\
		\ENSURE $\mathbf{X}_N, \mathbf{Y}_N$ as the normalized signals, $\bm{\mu},\bm{\sigma}$ as the mean, and standard deviation parameters  \\[5pt]
		\STATE $\bm{\mu}\leftarrow [\hspace{0.15cm}]$
		\STATE $\bm{\sigma}\leftarrow [\hspace{0.15cm}]$
		\FOR{$i$ in $1$ to $N_c$}
		\STATE $\mu_x \leftarrow \frac{1}{T}\sum_{j=1}^{T} \mathbf{X}(i,j)$
		\STATE $\sigma_x \leftarrow \sqrt{\frac{\sum_{j=1}^{T} (\mathbf{X}(i,j)-\mu_x)^2}{T}}$
		\STATE $\mathbf{X}_N(i,:) \leftarrow \frac{\mathbf{X}(i,:)-\mu_x}{\sigma_x}$
		\STATE $\bm{\mu}\leftarrow[\bm{\mu},\mu_x]$
		\STATE $\bm{\sigma}\leftarrow[\bm{\sigma},\sigma_x]$
		\ENDFOR
		\\
		\FOR{$i$ in $1$ to $N_E$}
		\STATE $\mu_y \leftarrow \frac{1}{T}\sum_{j=1}^{T} \mathbf{Y}(i,j)$
		\STATE $\sigma_y \leftarrow \sqrt{\frac{\sum_{j=1}^{T} (\mathbf{Y}(i,j)-\mu_y)^2}{T}}$
		\STATE $\mathbf{Y}_N(i,:) \leftarrow \frac{\mathbf{Y}(i,:)-\mu_y}{\sigma_y}$
		\STATE $\bm{\mu}\leftarrow[\bm{\mu},\mu_y]$
		\STATE $\bm{\sigma}\leftarrow[\bm{\sigma},\sigma_y]$
		\ENDFOR
		\\
		\RETURN $\mathbf{X}_N,\mathbf{Y}_N,\bm{\mu},\bm{\sigma}$
	\end{algorithmic}
\end{algorithm}

% \subsection{Real dataset}
% For the real dataset it is essential to select a package with a EOG reference electrodes for further performance measurments such as ME, MAE, MSE in EOG estimation stage. Because of that reason, we chose BCI competition IV, dataset 2a which relates to the motor imagination tasks \cite{tangermann2012review}. This dataset was recorded from nine healthy subjects in two different session (called Training, and Evaluation) with 288 trials for imagination of moving right, and left hands, feet, and tongue. In the beginning of each session, the subject is asked to open his/her eyes on the fixation cross for two minutes, closing the eyes for one minute, and finally moving the eyes for one minute. This procedure can help the researchers remove the EOG artifact. Also, the EEG is recorded in twenty two channels in 10-20 international standards, and three monopolar EOG references for recording the EOG signals (shown in Fig.\ref{fig:eogfig}). The sampling frequency is 250 Hz and the signals are filtered in range of 0.5 Hz to 100 Hz, and a 50 Hz notch filter was applied to remove the power supply noise. 
% % TODO: \usepackage{graphicx} required
% \begin{figure}
% 	\centering
% 	\includegraphics[width=0.3\linewidth]{eogfig.eps}
% 	\caption{The placement of EOG electrodes for BCI competition IV- dataset 2a}
% 	\label{fig:eogfig}
% \end{figure}
\section{Results}\label{results}
In this section, various metrics are evaluated for single and multi channel EOG artifact removal from EEG recordings.

\begin{figure}
	\centering
	\includegraphics[width=\linewidth]{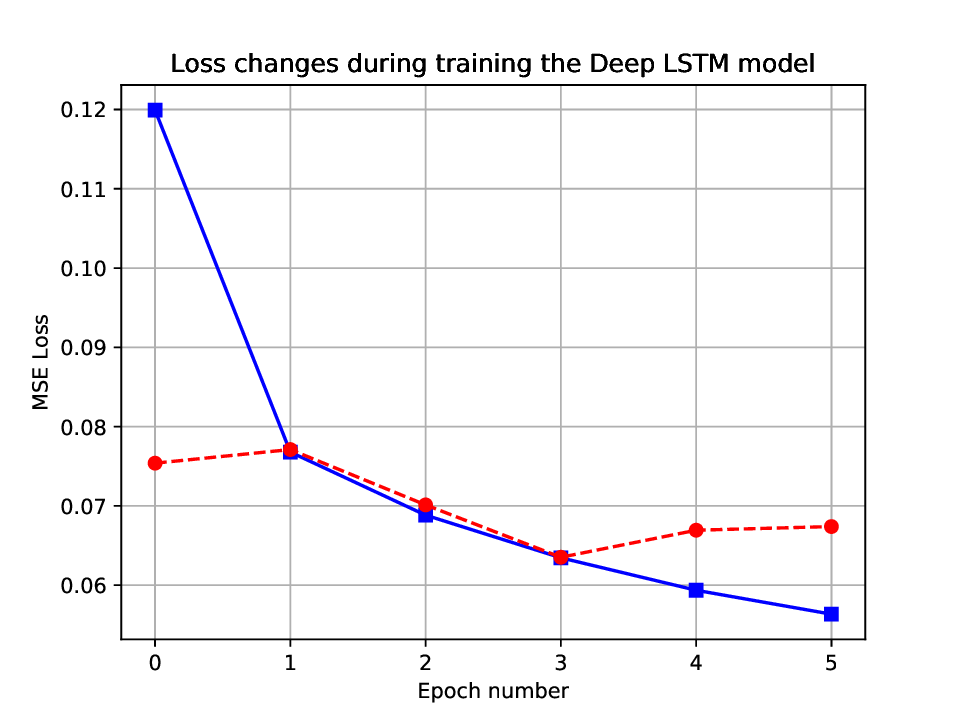}
	\caption{Changes in MSE Loss During Training Scheme. The red plot represents the validation set, while the blue plot corresponds to the training data. Epoch 0 denotes the first epoch.}
	\label{fig:training}
\end{figure}

\begin{figure}[!ht]
	\centering
	\includegraphics[width=\textwidth]{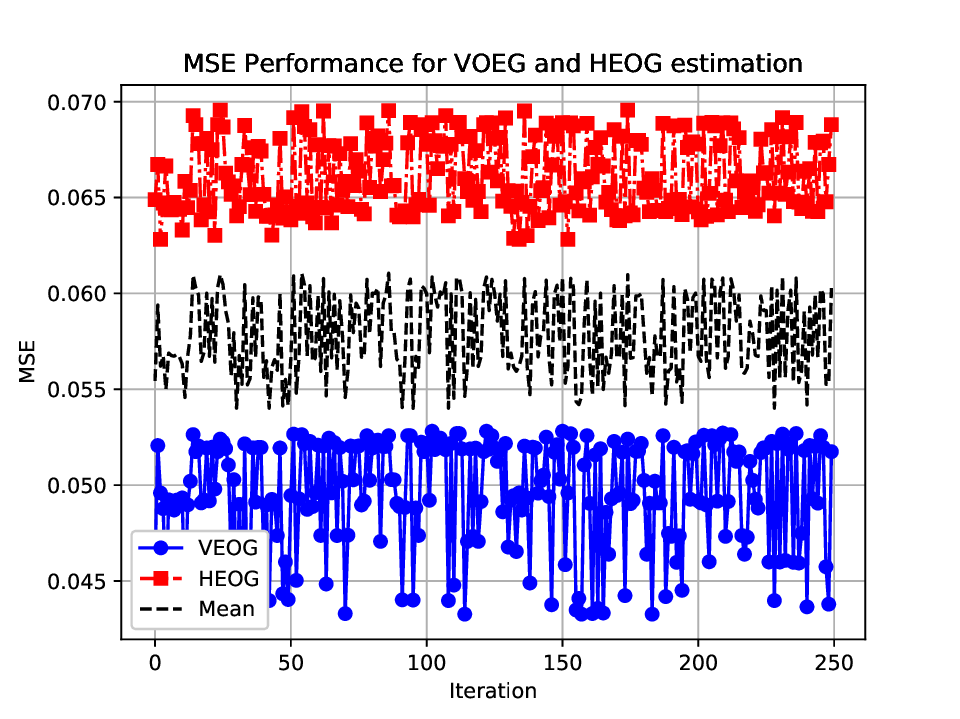}
	\caption{The performance of VEOG and HEOG estimation over 250 iterations.}
	\label{fig:mseklados}
\end{figure}

\begin{figure}[ht!]
	\centering
	\subfloat[HEOG]{\includegraphics[width=\textwidth]{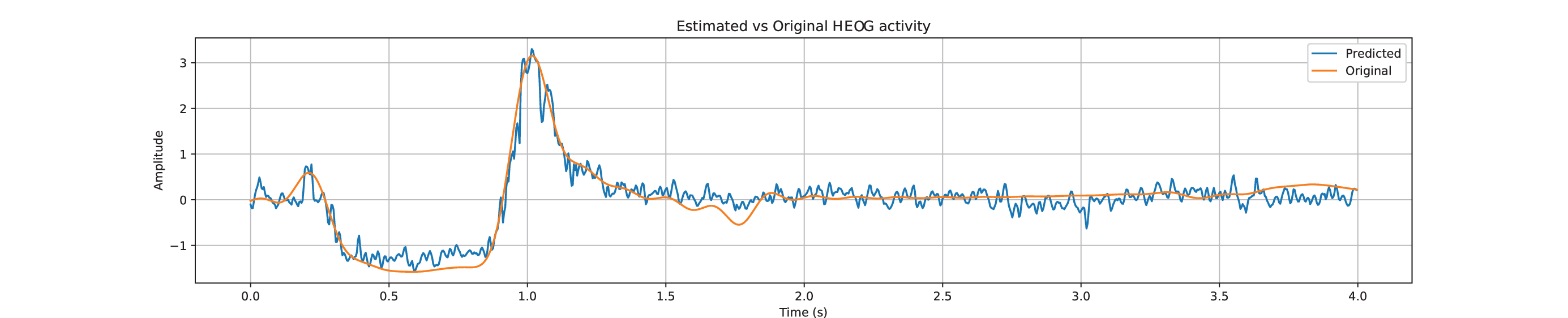}}\\
	\subfloat[VEOG]{\includegraphics[width=\textwidth]{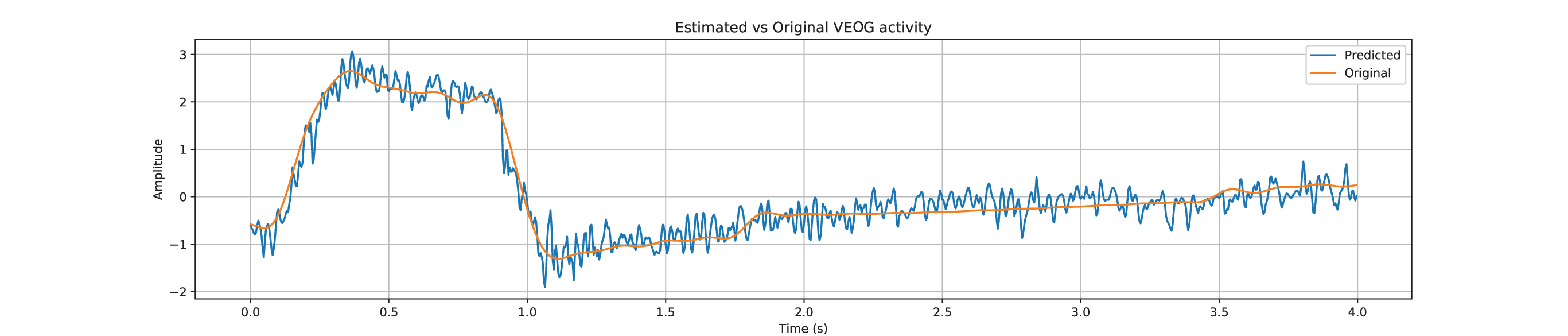}}
	\caption{Comparison between estimated EOG and its original waveform in a sample segment.}
	\label{fig1}
\end{figure}

\begin{figure}[!ht]
	\centering
	\includegraphics[width=\textwidth]{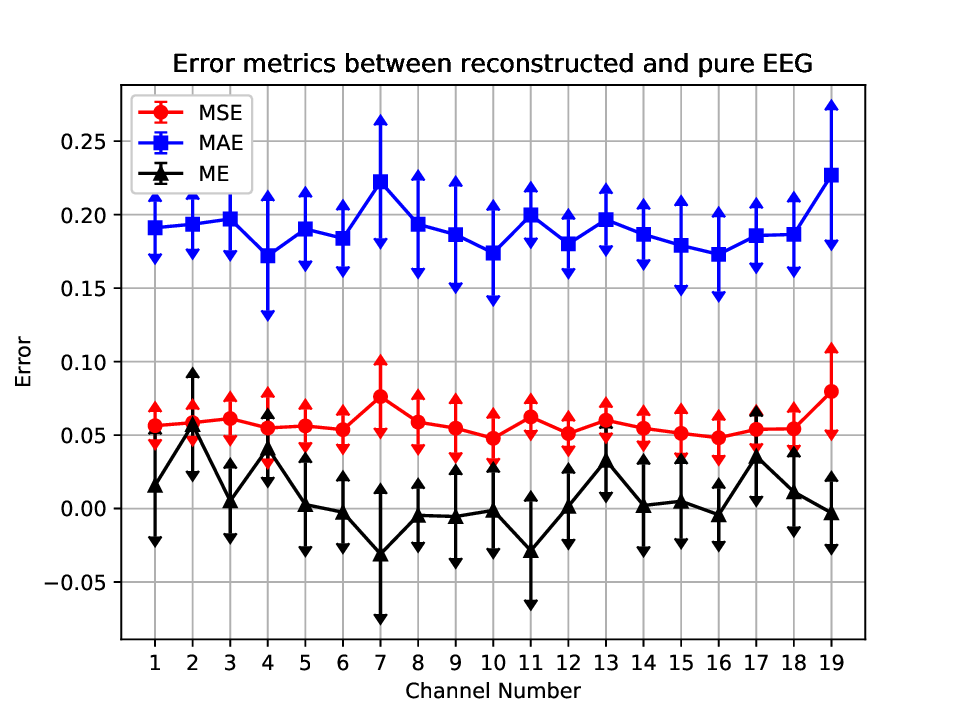}
	\caption{Multi-channel EEG de-artifactualization performance with proposed method. The arrows represent the range of standard deviation.}
	\label{fig:errorsklados}
\end{figure}

The entire procedure is divided into three stages: 1) Training the Deep LSTM network using contaminated EEG recordings and EOG reference channels with training segment samples. 2) Estimating EOG recordings for the testing data. 3) Combining EEG recordings with the estimated EOG activity, performing ICA, and reconstructing the artifact-free recording. Since both clean and contaminated recordings are provided in the dataset, we can measure the performance of both EOG estimation and EEG denoising using the proposed algorithm and compare it with previous literature. The overall network architecture for this dataset consists of nineteen input channels and two output channels for EOG estimation. The MSE loss values during the 50 epochs of training are depicted in Figure \ref{fig:training}. It is important to note that training was stopped early due to early stopping criteria with a patience of two epochs.
At this stage, we can obtain the EOG estimations from the trained Deep LSTM network. To evaluate the performance, we randomly selected 30
\% of the semi-simulated datasets using a cross-subject paradigm. This ensures a fair evaluation, as training on segments from one subject and testing on segments from the same subject could introduce bias. The remaining test data was divided into 250 different segments with variable lengths and overlapping samples. The results are presented in Figure \ref{fig:mseklados} and Table \ref{tab1}. Additionally, for a visual comparison, a sample output of the Deep LSTM network for both HEOG and VEOG estimation is shown in Figure \ref{fig1}, alongside the original EOG activation in the dataset. 
% TODO: \usepackage{graphicx} required
\begin{table}[!ht]
	\centering
	\caption{Quantified EOG estimation performance on test set over 250 iterations.}
 \label{tab1}
	\begin{tabular}{c|c}
		\hline
    \hline
		EOG channel	&  Error (Mean $\pm$ STD)\\
		\hline
		VEOG & $0.051\pm0.011$  \\
		\hline
		HEOG & $0.040\pm 0.007$ \\
		\hline
		\textbf{Average}	&  $\mathbf{0.046\pm 0.005}$\\
		\hline
    \hline
	\end{tabular}
\end{table}
To perform the final analysis, we evaluate the cleaning of the artifactual EEG recordings on the test data, which was not seen during training. The estimated EOG signals are concatenated with the observed contaminated data, and then the FastICA algorithm is applied to extract the artifactual sources and clean the contaminated EEG signals. For each of the nineteen channels, we compare the estimated clean EEG with the pure signals provided in the semi-simulated dataset. The results for each channel are shown in Figure \ref{fig:errorsklados}.

\begin{figure}[!ht]
	\centering
	\includegraphics[width=1\textwidth]{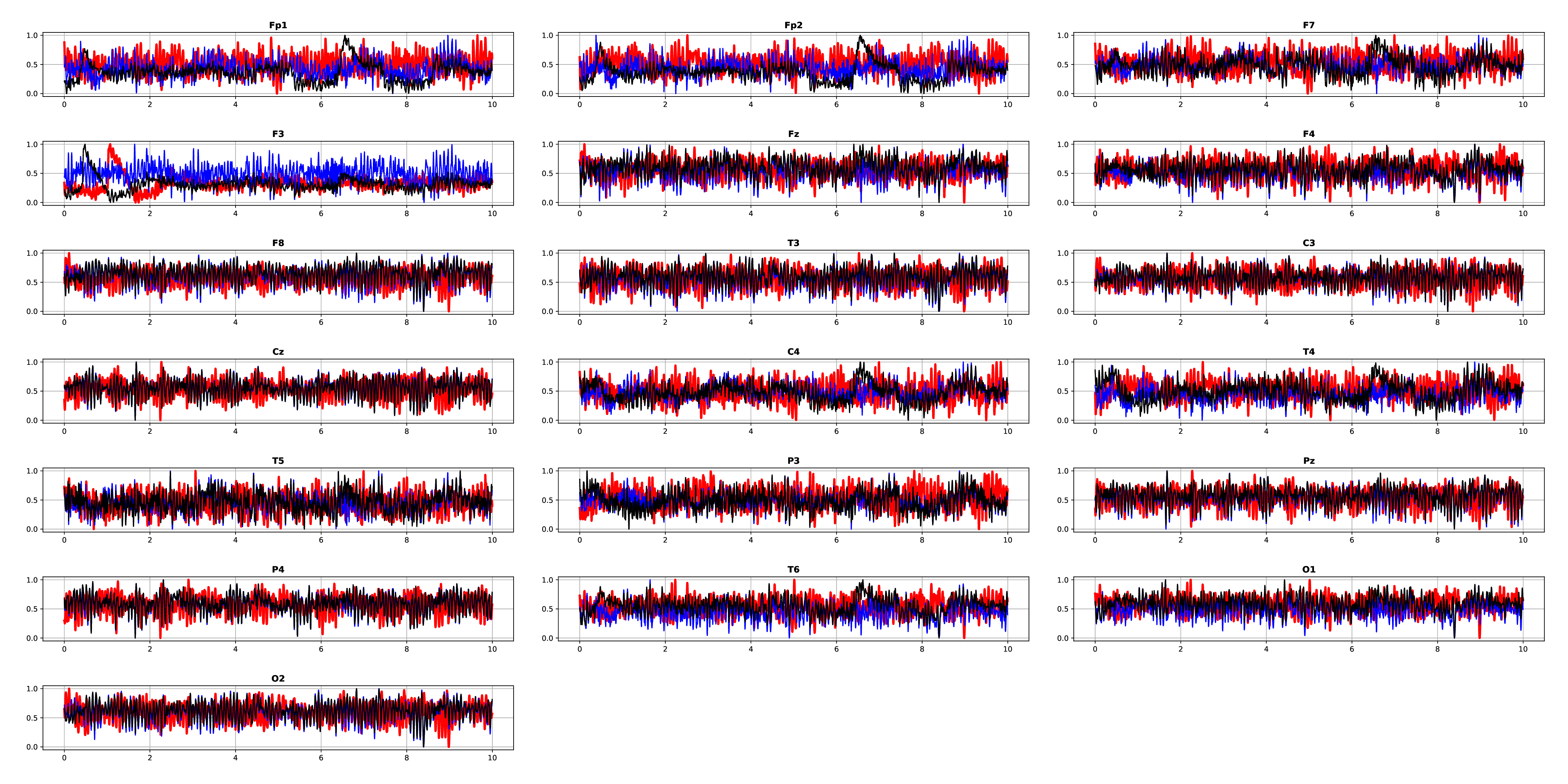}
	\caption{Sample of Cleaned EEG Using the Proposed Method: Zoom-In for Details. The red curves represent the original pure EEG, while the blue curves depict the output of the proposed algorithm.}
	\label{fig:sampleklados}
\end{figure}

In order to visualize the time series data, Figure \ref{fig:sampleklados} presents the reconstructed EEG signals for all nineteen channels alongside the corresponding pure EEG signals. To facilitate comparison and eliminate the effect of scale uncertainty, the signals are min-max normalized between 0 and 1 in each channel before plotting.
% TODO: \usepackage{graphicx} required
% TODO: \usepackage{graphicx} required
The results clearly indicate that the overall error between the reconstructed clean EEG and the pure EEG is relatively low (approximately 0.05 MSE, 0.16 MAE, and 0.02 ME) across all channels in the main dataset. To further evaluate the effectiveness of our proposed method, we compared it with two recent deep learning-based approaches: SAE-RLS by \cite{yang2016removal}, and DLN-SAE by \cite{yang2018automatic}. The comparison results are presented in Table \ref{tab:comparison}, highlighting the superior performance of our suggested method over the other deep learning-based approaches.

\begin{table}[!ht]
	\centering
	\caption{Comparison of the proposed method with the existing similar literature. The reported values are average errors between estimated artifact-free EEG and the original EEG recordings. }
 \label{tab:comparison}
	\begin{tabular}{c|c|c|c}
		\hline
    \hline
		 Method &  MSE & MAE & ME\\
		\hline
		SAE-RLS \cite{yang2016removal} & $0.15$ & $0.39$ & $\mathbf{0}$ \\
		\hline
		DLN-SAE \cite{yang2018automatic} & $0.08$ & $0.29$ & $0.01$ \\
  \hline
  \textbf{Proposed method} & $\mathbf{0.05}$ & $\mathbf{0.16}$ & $0.02$\\
		\hline
  \hline
	\end{tabular}
\end{table}

% TODO: \usepackage{graphicx} required
\begin{figure}[!ht]
	\centering
	\includegraphics[width = \textwidth]{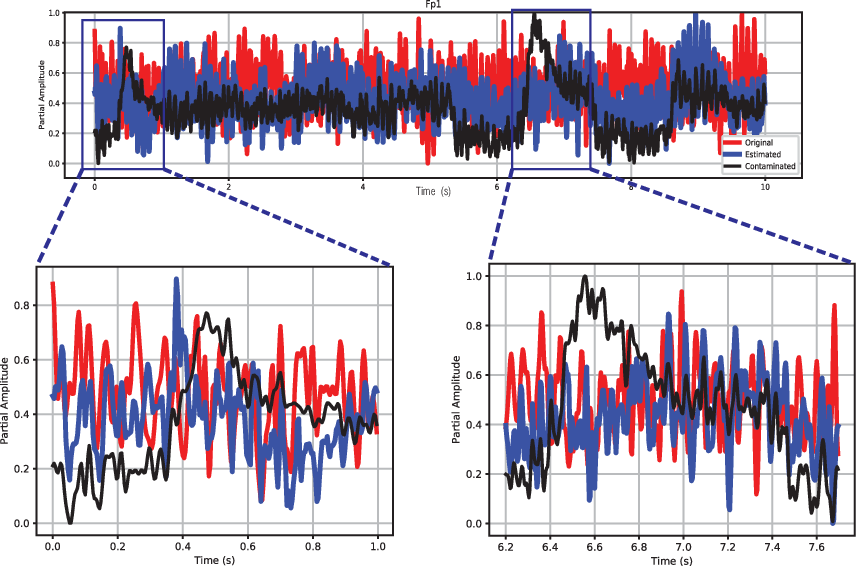}
	\caption{Detailed visualization of the performance of the proposed algorithm in Fp1 channel.}
	\label{fig:fp1fig}
\end{figure}

\section{Discussion}\label{discussion}
This paper presents a novel method for estimating EOG signals using a deep LSTM network and subsequently removing EOG artifacts from contaminated EEG recordings through ICA. The results in Figures \ref{fig:sampleklados} and \ref{fig:fp1fig} demonstrate that the proposed algorithm performs effectively even in frontal channels such as Fp1, successfully detecting EOG peaks in EEG recordings and eliminating their effects using ICA. This is attributed to the strong performance of ICA in removing EOG artifacts and the potential of LSTM networks in various EEG-related tasks \cite{wang2018lstm,nagabushanam2020eeg,zhang2019classification,lee2023lstm}, indicating their ability to learn underlying relationships in EEG data.

The results in Table \ref{tab:comparison} indicate that the proposed method achieves superior performance in terms of MSE and MAE compared to the DLN-SAE method \cite{yang2018automatic}, both of which are metrics that are unaffected by the sign of the EOG amplitude. However, for the ME metric, which is highly sensitive to the polarity of the EOG amplitude, the DLN-SAE method demonstrates better performance. The enhanced performance in terms of MSE and MAE can be attributed to the utilization of a more intricate network architecture, specifically LSTM, in contrast to the combination of MLP and SAE, as previously shown by \cite{li2022deep}.

While the proposed pipeline primarily focuses on EOG artifact removal from EEG signals, it can also be employed for EOG estimation in applications such as gaze detection using EEG signals when EOG recordings are not available.

However, it is important to acknowledge certain limitations of this approach. Firstly, the fine-tuning and adjustment of the LSTM network were carried out through trial and error, suggesting the need for comprehensive approaches to optimize hyperparameters, including the number of hidden layers and LSTM units, in different applications and datasets. Secondly, the availability of a gold standard non-contaminated dataset, as demonstrated in \cite{klados2016semi}, was crucial for this study. Therefore, further investigations are required to ensure the applicability of this approach in real-world scenarios. For instance, transfer learning could be explored to train models on one dataset and evaluate their performance on other datasets. Given the high subject-to-subject variability, the generalizability of the approach remains a key question to be addressed in future research. Lastly, while FastICA was utilized as the core method in the ICA stage of this paper, other BSS techniques may outperform FastICA in different applications. Exploring alternative BSS methods could be an avenue for future directions in this study.

These limitations highlight important aspects for future research in order to address them and further enhance the efficacy and applicability of the proposed method.

\section{Conclusion}\label{conclusion}
In this paper, we introduced an LSTM-ICA methodology for effectively removing EOG artifacts from EEG recordings in the absence of external EOG recordings. The proposed approach was evaluated using a dataset that included both contaminated and non-contaminated EEG recordings. The performance of the model was assessed in two scenarios: single-channel and multi-channel EEG data. Our methodology demonstrated superior performance compared to existing deep learning-based approaches, emphasizing the potential of combining ICA and LSTM in future EEG studies.

\section*{Conflict of Interest} The authors have no conflicts of interest to disclose.
\section*{Ethics statement} This research did not involve the collection of any data from human subjects.
\section*{Funding statement} This research was not supported by any funding resources. 

\bibliography{elsarticle-template.bib}
%%%%%%%%%%%%%%%%%%%%%%%%%%%%%%%%%%%%%%%%%% Revised
\end{document}